\documentclass[journal]{vgtc}                

\usepackage{mathptmx}
\usepackage{wrapfig}
\usepackage{times}
\usepackage{color}
\usepackage[table,xcdraw]{xcolor}
\usepackage[T1]{fontenc}
\usepackage{mathptmx}
\usepackage{amsmath}
\usepackage[pdftex]{hyperref}
\usepackage{booktabs}
\usepackage{textcomp}
\usepackage[normalem]{ulem}
\usepackage{ccicons}
\usepackage{todonotes}
\usepackage{lettrine}
\usepackage{enumitem}
\usepackage{multimedia}
\usepackage{media9}
\usepackage{subfig}
\usepackage{float}

\onlineid{0}

\vgtccategory{Research}

\vgtcinsertpkg

\title{Why Authors Don't Visualize Uncertainty}

\author{Jessica Hullman}
 \authorfooter{
\item
 Jessica Hullman is with Northwestern University. E-mail: jhullman@northwestern.edu.

}

\abstract{Clear presentation of uncertainty is an exception rather than rule in media articles, data-driven reports, and consumer applications, despite proposed techniques for communicating  sources of uncertainty in data. 
This work considers, Why do so many visualization authors choose not to visualize uncertainty? I contribute a detailed characterization of practices, associations, and attitudes related to uncertainty communication among visualization authors, derived from the results of surveying 90 authors who regularly create visualizations for others as part of their work, and interviewing thirteen influential visualization designers. 
My results highlight challenges that authors face and expose assumptions and inconsistencies in beliefs about the role of uncertainty in visualization. In particular, a clear contradiction arises between authors' acknowledgment of the value of depicting uncertainty and the norm of omitting direct depiction of uncertainty. To help explain this contradiction, I present a rhetorical model of uncertainty omission in visualization-based communication. I also adapt a formal statistical model of how viewers judge the strength of a signal in a visualization to visualization-based communication, to argue that uncertainty communication necessarily reduces degrees of freedom in viewers' statistical inferences. 
I conclude with recommendations for how visualization research on uncertainty communication could better serve practitioners' current needs and values while deepening understanding of assumptions that reinforce uncertainty omission. 
} 


\CCScatlist{ 
 \CCScat{K.6.1}{Management of Computing and Information Systems}%
{Project and People Management}{Life Cycle};
 \CCScat{K.7.m}{The Computing Profession}{Miscellaneous}{Ethics}
}

\begin{document}



\maketitle

\section{Introduction}
Consider the last few visualizations you encountered outside of scientific publications. Perhaps you read a news article backed by data, or researched a product, or used a transit application to plan a trip. Did the visualizations depict uncertainty--the possibility that the observed data or model predictions could take on a set of possible values?

Chances are, they did not. 
Of 612 data visualizations from 121 articles published online in February 2019 by a set of leading purveyors of data journalism, social science surveys, and economic estimates\footnote{Sample includes: The Bureau of Labor Statistics, The Economist Graphic Detail, FiveThirtyEight, the OECD, the Pew Foundation, the Urban Institute.},
449 (73\%) presented data intended for inference,
but only 14 (3\%) portrayed uncertainty visually, either by depicting explicit quantifications like intervals or conveying variance through raw data.


There are many reasons why visualization practitioners and other data communicators might choose to exclude uncertainty information. Some stem from user experience concerns: authors may perceive uncertainty to be challenging to viewers based on psychological complexity. When a probability is 0 or 1, it can be validated; when it is not, how it should be interpreted is debatable~\cite{definetti1977}.
When visualizations are used to communicate and inform, as in public-facing reports, media presentations, business reporting, and end-user applications, the attention of the viewer is often at a high premium. 
Authors may question the necessity of adding more information to an already complex display. 
Moreover, even expert scientists and analysts can struggle to accurately calculate uncertainty~\cite{webster2003communicating}, and finding visualization approaches that are accessible and applicable across data and visual formats remains a research challenge. 
Still other reasons may be rhetorical. Authors may want to signal confidence, 
perceiving uncertainty to undercut the credibility of their results based on unstated norms. 

The goal of this paper is to understand: Why isn't uncertainty visualized more often in visualization-based communication?  
The first step in this inquiry is to identify how visualization authors perceive and use uncertainty representation. I contribute the results of a survey of 90 professional visualization authors and interviews with thirteen influential professional visualization designers and journalists. 
I use these results to characterize how communicative visualization authors understand and engage with uncertainty visualization. In the process, I identify unaddressed challenges in current practice that visualization researchers may find worth devoting attention to. For example, authors express the need for resources to help them develop text-based explanations and produce broadly understandable visual representations. 

My second contribution is a rhetorical model of uncertainty omission in communicative visualization. A set of \textit{model tenets} summarize beliefs that authors appeal to in rationalizing the omission of uncertainty from visualizations. The model describes visualization as a means of both conveying and producing ``signal'', which is validated by the author's analysis process for both authors and viewers, but believed to be obfuscated by exposing uncertainty. 

The goals are twofold. First, by capturing pervasive beliefs in three model tenets, the model provides a concise explanation for how a norm of uncertainty omission may persist despite many authors' professed belief in the importance of uncertainty visualization. Secondly, 
an understanding of pervasive rationales can help visualization researchers and others devote their efforts toward addressing the beliefs or needs for resources that motivate well-intentioned authors to omit uncertainty. 

My third contribution is to propose a formalism for the signal-judgment mechanism implied by the rhetorical model. I apply a statistical model of graphical inference proposed for analysis to communicative visualization, using it to demonstrate how uncertainty visualization necessarily reduces flexibility in viewers' processes for judging a visualization's signal. I summarize multiple logical inconsistencies implied by authors' perceptions as summarized in the rhetorical model. 
I conclude with a discussion of 
how visualization research could more directly
address authors' needs and deepen understanding of perceptions around uncertainty.

\section{Background}
\subsection{Communicating Uncertainty for Decision Making}
Scientific discourse, which lays the groundwork for evidence-based decision making, is largely a discussion of uncertainty. However, often the level of detail in the scientific literature on a topic or method is inappropriate for a decision-maker, either because it is too minute to easily decipher or fails to state assumptions that are important but assumed known among scientists~\cite{fischhoff2014communicating}. 
An analogous scenario occurs in data analysis contexts, where analysts spend time identifying and evaluating the robustness of patterns in visualized data. Reproducing their complete analysis workflow to communicate results would be overwhelming to users in many cases. Hence, visualization authors engage in a rhetorical process involving editorial decisions to omit, emphasize, and otherwise transform results to guide an audience toward intended interpretations~\cite{hullman2011}.


For a visualization author to successfully communicate uncertainty, 
they must 1) recognize the value of uncertainty information for receivers of their messages, and 2) identify effective ways to communicate it. Threats to the successful completion of both steps can arise. 

For example, an author may choose not to acknowledge uncertainty if they feel uncomfortable conveying numerical estimates of risk. Prior research suggests that experts or analysts typically prefer to communicate uncertainty in qualitative terms, fearing misinterpretation if they state uncertainty, though decision makers and other end-users of reports typically prefer to receive uncertainty in precise quantitative terms~\cite{erev1990verbal,olson1997patterns}. 
Other proposed reasons that a communicator may be resistant to the idea of presenting uncertainty include: fear that uncertainty information will imply unwarranted precision in estimates~\cite{fischhoff2012communicating}, a tendency to think that the presence of uncertainty is common knowledge~\cite{fischhoff2012communicating}, a tendency to think that non-expert audiences will not understand the uncertainty information~\cite{fischhoff2012communicating,manski2018lure}, a belief that not presenting uncertainty will simplify decision making~\cite{manski2018lure}, a belief that people cannot tolerate uncertainty~\cite{manski2018lure}, a belief that uncertainty creates negative feelings~\cite{manski2018lure}, a belief that not presenting uncertainty will make it easier to coordinate beliefs~\cite{manski2018lure}, and a belief that presenting it will make their message seem less credible~\cite{fischhoff2012communicating,manski2018lure}. 



Breakdowns may also occur when uncertainty is communicated. 
Fischoff~\cite{fischhoff1995risk} summarizes possible breakdowns in a set of ``developmental stages'' in risk communication: ``All we have to do is: 1) get the numbers right; 2) tell them the numbers; 3) explain what we mean by the numbers; 4) show them they've accepted similar risks in the past; 5) show them it's a good deal for them; 6) treat them nice; 7) make them partners; 8) all of the above.'' 

Scholarly work describing rationales like those above has largely been based on empirical evidence around uncertainty comprehension (e.g., the large body of research on overconfidence among those presented with uncertain information~\cite{fischhoff1977knowing,fischhoff1982subjective,lichtenstein1981calibration}) or anecdotal evidence (e.g., a former director of the Congressional Budget Office's (CBO) statement that ``You can't give the client a bound. The client needs a point''~\cite{manski2018lure}). Researchers in economics and visualization have suggested that omitting uncertainty information from data presentations may result from an unstated norm~\cite{finger2002,manski2015communicating}.
I attempt to directly elicit authors' perceptions about uncertainty and associated norms by conducting a survey and interviews.  


Some studies have examined how data workers manage uncertainty--including attempts to understand, diagnose, minimize, exploit, and suppress it--as they attempt to extract insights from data~\cite{boukhelifa2017data,kale2019decision,lipschitz1997,schunn2012psychology,skeels2010}. 
These accounts establish that even ``experts'' struggle to properly quantify and manage uncertainty. Several studies describe data workers' perceptions of the difficulty of communicating uncertainty and a tension between their goals of transparency and accuracy and decision-makers' desires for simplicity ~\cite{boukhelifa2017data,kale2019decision,skeels2010}.

\subsection{Visualizing Uncertainty}
Research in uncertainty visualization starts with quantified uncertainty and considers how to best visually express it. Researchers have taxonomized sources of, and representations for, uncertainty~\cite{griethe2006,johnson2003,maceachren1992visualizing,pang1997,potter2012,skeels2010,thomson2005} and demonstrated techniques intended for analysts and other experts (e.g.,~\cite{bastin2002visualizing,davis1997modelling,ehlschlaeger1997visualizing,potter2012,wickham2010,wittenbrink1996glyphs}). More recently, researchers have proposed and evaluated visualizations intended for non-experts as well as experts, such as hypothetical outcome plots~\cite{hullman2015,kale2018hypothetical}, quantile dotplots~\cite{fernandes2018uncertainty,kay2016ish}, value suppressing palettes~\cite{correll2018}, and gradient and violin plots~\cite{correll2014}). 
However, few researchers in uncertainty visualization have devoted attention to why uncertainty visualization remains uncommon in practice, or whether authors' needs are met by the outputs of research on visualizing uncertainty. 
Exceptions include papers that speculate on how authors may see uncertainty metadata as no different from other metadata and include it only as secondary~\cite{boukhelifa2009}, may struggle to find representations that are compatible with the data, tasks, and organizational goals~\cite{boukhelifa2009}, and may find it difficult to evaluate the impact of uncertainty on visualization-based judgments~\cite{boukhelifa2009,hullman2019}. 



\section{Methods}
The characterization of practices and associations around uncertainty visualization that I contribute is based on formative research conducted over seven months in 2018 and 2019. During this time, I surveyed visualization authors and interviewed a smaller set of authors I perceive to be influential among visualization practitioners. 
The questions posed in this work are unlikely to be answered to satisfaction by a single study; however, for the purposes of this initial inquiry I stopped aggressively recruiting new participants once responses seemed to converge on a set of widely held rationales.

\subsection{Survey of Visualization Creators}
Between December 2018 and May 2019 I conducted an online survey of 90 visualization authors in a convenience sample. I shared the survey link on Twitter several times over this period. The recruitment text described target respondents as those who ``regularly create visualizations for others.'' The headline began ``Attn vis designers, developers, journalists...'' To incentivize participation, one \$50 Amazon gift card was awarded to a randomly drawn respondent for each 20 respondents (5 gift cards total). 
\begin{figure}[htb]
 \centering
  \includegraphics[width=\columnwidth]{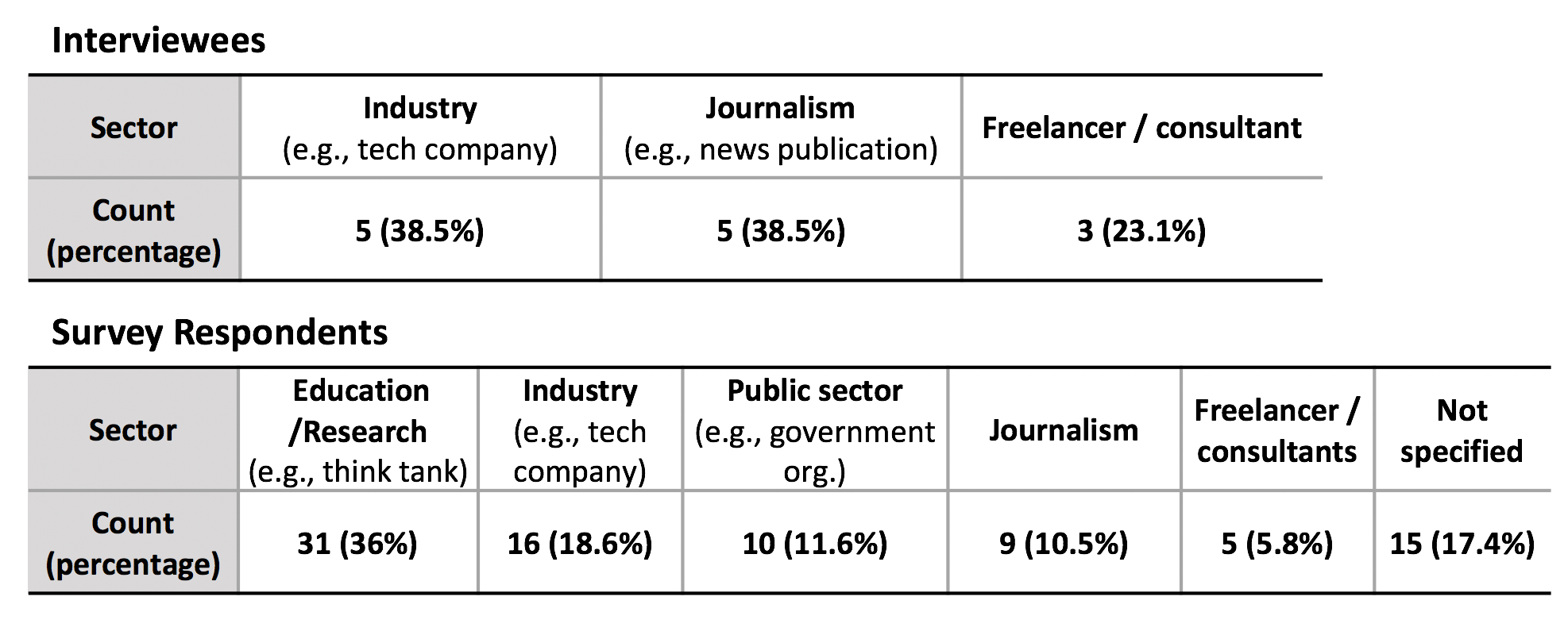}
  \vspace{-0.25in}
  \caption{Summary of the types of organizations interviewees and survey respondents are associated with.}
  \label{fig:participants_sector}
\end{figure}


\subsubsection{Survey Design}
The online survey (17 required questions) aimed to identify how often respondents depict uncertainty, why they sometimes choose not to, and what sorts of design guidelines or other considerations they associate with uncertainty. Six required questions in the first section of the survey asked respondents to consider their visualization work over the past year and state how often, and how, they had visualized uncertainty, and whether they had tried to include it but opted not to, and why not. A second section of 8 required questions asked respondents to rate their agreement with the statements depicted in Figure~\ref{fig:survey_result} using a 5 point scale labeled from Strongly Disagree (1) to Strongly Agree (5). In a final section of 3 required questions, respondents were asked to describe their position, the audience for which they most frequently created visualizations, the frequency with which they authored visualizations for others, and optionally the organization they worked for and links to their work.


\subsubsection{Participants}
A total of 90 visualization authors completed the survey as of May 2019. Table 1 describes the sample. 
Given the use of a convenience sample recruited on social media, selection bias is likely, such as towards authors who have spent time thinking about uncertainty or are more likely to think it should be expressed. 
I did my best to remain aware of this possible selection bias in analysis by avoiding interpreting the absolute frequency of responses or assuming the rationales provide an exhaustive description. 

\subsection{Interviews with Influential Visualization Creators}
In gain a deeper understanding of
potential causes of inconsistencies in attitudes and practices that the survey results exposed,
I conducted semi-structured interviews with thirteen people I consider to be ``visualization influencers'' who do not self-identify as researchers (6 female) between February and June 2019. Figure~\ref{fig:participants_sector} summarizes the sectors in which interviewees worked. 
All thirteen interviewees have Twitter accounts, with a median of 4.7k followers (mean: 11k). 

\begin{wrapfigure}[16]{r}{0.26\textwidth} 
\vspace{-0.25in}
  \begin{center}
    \includegraphics[width=0.26\textwidth]{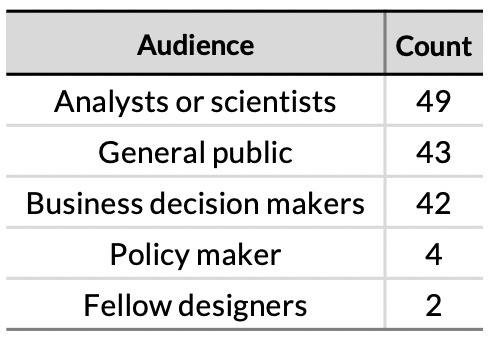}
   \end{center}
      \caption{The list of audiences that survey respondents described creating visualizations for. Respondents could choose multiple answers.}
  \label{fig:audience}
\end{wrapfigure}

\subsection{Analyzing Responses}
All survey responses were recorded via Google forms. Eleven of the interviews were transcribed by a professional transcriptionist. Two of the interviews were not recorded. 

To identify values and rationales in the results of the above activities, I started with open coding to identify themes, then iterated over the responses several times until the codes stabilized~\cite{creswell2007}. For both survey and interview results, I categorized understandings of the term ``uncertainty visualization,'' descriptions of how participants depicted uncertainty (if applicable), rationales for not expressing uncertainty, and perceptions of the value of uncertainty. Related to rationales for not expressing uncertainty, I also coded statements that seemed aimed at describing a perceived status quo, which became the basis for the rhetorical model. 

\section{Summary of Uncertainty Associations \& Practices}

\subsection{Uncertainty Visualization: What and How?}
To set the context for the online survey and interviews, participants were first asked to describe what they thought of when hearing the term ``uncertainty visualization.'' The most prevalent responses described uncertainty as an interval, range, or region (roughly half of respondents), often in conjunction with error bars. Mentions of possible outcomes and/or the possibility that plotted values may not represent the real values (roughly 15\%) were also common. Probability, confidence, variance, error, missing data, and sampling and modeling artifacts were each mentioned by multiple respondents. While most of these initial impressions did not include value judgments, a handful of online survey respondents defined uncertainty visualization by referring to tensions between uncertainty and other design goals, like ``how to convey the possibility of divergence from a projection without casting doubt on the fundamental measurements or reports'' and ``a visualization that doesn't have a clear message.'' 

When asked whether they had depicted uncertainty in any visualizations they had created in the last year,
roughly three quarters (76\%) of survey respondents replied yes. 
When asked about occasions on which authors had communicated uncertainty, survey respondents and interviewees mentioned a number of techniques consistent with an understanding of uncertainty as error or probability. Surveyed authors most commonly described using interval representations (75\%), visual variables to encode probability or confidence (75\%), density plots or histograms (54\%), and text representations (51\%).
However, uncertainty as a \textbf{qualitative expression of a gap in knowledge} came up in most interviews with interviewees as well as several survey responses. 62\% of survey respondents had used text to warn their viewers of the potential for uncertainty in results. Several interviewees described how they had more experience thinking deeply about qualitative forms of uncertainty (I1, I4, I5), like a lack of available explanation for how a particular quantitative prediction would come to be (e.g., in what order and at how constant a rate certain climate changes come about). 

Several respondents spontaneously associated uncertainty with \textbf{process}, including ``Even communicating uncertainty may also require communicating fundamental hermeneutics and methodologies of that information,'' and 
``the answer to the question someone will inevitably ask `how did you get these numbers?'''

Uncertainty was also associated with \textbf{resolution}. One interviewee who works in industry described uncertainty as a symptom of the granularity or resolution of a plot: ``I approach uncertainty a lot of times by trying to peel back layers instead of being like `Let's look at an average value and try to get some sense of what the distribution is under that. Can I actually just show all of the data that they have available?'' (I6). Roughly 5\% of survey respondents described uncertainty as variance, which is shown by simply plotting raw data.

Multiple survey respondents and interviewees alluded to uncertainty visualizations that defied the conventional definition of uncertainty as a quantitative probabilistic representation. Uncertainty was
associated with \textbf{visual inexactitude}, or encodings intentionally chosen to imply imprecision, by four interviewees and a handful of survey respondents. One freelance interviewee described his choices in visualizing imprecise estimates: 
``What we did with this one, because anything that would have give us a precise treatment would've been making something precise that is not, we actually used different bins.... Because it's not plotting on an exact axis point, risk scale equals implied precision. That was the intent'' (I4). The interviewee went on to mention perceptual accuracy rankings for visual encodings ``if you back down to the classic perceptual rankings, that kind of thing, those that are lower down the rankings of perceptual accuracy do work in my view.'' Another freelance interviewee (I1) described the use of circular area and conjoined marks with difficult to resolve boundaries as strategies for visualizing uncertainty that did not involve explicitly mapping quantified uncertainty to visual properties. 


Several survey respondents and four interviewees (I1, I4, I6, I8) also described the possibility of communicating uncertainty through a \textbf{progression of views depicting patterns that are more to less probable}, an interpretation that is not usually considered in scholarly research on uncertainty visualization. 
A survey respondent attributed a design pattern to the work of one of the interviewees: ``[interviewee] shows a useful way of communicating uncertainty with small multiples. It breaks down observations into three bins: 1) this is something that we've seen at least once; 2) this is something that we see sometimes; and 3) this is something that we see all the time.'' 
To respondents, patterns like these appeared to inhabit a middle ground between qualitative uncertainty that might be best expressed in text and explicit visual representation of quantified uncertainty. 


\subsection{Failure \& Voluntary Omission} 
Though the majority of online survey respondents and interviewees said they had communicated uncertainty in at least one visualization in the previous year, their estimated percentage indicated that uncertainty visualization is a relatively rare phenomena: over one third of online survey respondents indicated that 10\% or less of the visualizations they had created communicated uncertainty. Only one quarter described communicating uncertainty in 50\% or more of their visualizations.

Moreover, nearly half of survey respondents admitted to having intended or attempted to communicate uncertainty in a visualization they had created in the prior year, but had not ultimately included the uncertainty. Roughly one third of the interviewees described instances where uncertainty information was calculated or discussed among the authoring team but not included. 
Frequently cited reasons among survey respondents for not including uncertainty on these occasions were not wanting to confuse or overwhelm viewers (53/85; 62\%), not having access to uncertainty information (40/85; 47\%), not knowing how to calculate uncertainty (22/85; 26\%), and not wanting to make data seem questionable (15/85; 17\%).

\subsubsection{Expected Difficulties for Viewers}  
Concerns that uncertainty information may overwhelm or confuse viewers were echoed in other comments provided by survey respondents and interviewees. 
The belief that a visualization that explicitly represented uncertainty requires more effort than a visualization without it was implicit in many of these comments. 
One freelance interviewee described how the efforts required in readers' time and energy and sophistication should be ``rewarded with sufficient insight to understand [the uncertainty information]'' (I4). Another who works in industry described how explicit uncertainty representation means that readers ``have to come to you with patience, already have the patience needed to endure your explanation'' (I8). 
That uncertainty could provoke psychological anxiety was also suggested by several interviewees; as one journalism interviewee put it ``People have a lot of anxiety once they acknowledge even a small gap in something'' (I3). 
Survey respondents and several interviewees (I2, I8) alluded to empathy for the viewer as a possible motivation for uncertainty omission: ``Because I'm separated from the analysis, my question often ends up in rhetoric and ethics a lot more than it does `Am I truthfully portraying information?' It's much more about what does the audience care about, building a sort of empathy, if I can, with the audience; what difficulties they're going to encounter trying to learn the information'' (I8).

\subsubsection{Difficulties for Authors}
In contrast to viewer-related concerns, not having access to uncertainty information or not knowing how to calculate uncertainty are \textbf{author resource limits}. Their prevalence suggest that some authors do not have the tools or knowledge needed to follow through with uncertainty representation.
When asked for additional reasons that they had not presented uncertainty information in their recent visualizations, a handful of survey respondents mentioned time and other constraints directly. Multiple authors also referred to not being able to visualize uncertainty satisfactorily. As one visualization interviewee from industry described, ``we lack canonical forms for lay audiences'' (I7).

Several other interviewees described how stakeholders they worked with on visualizations questioned the value of the effort required to include uncertainty explicitly: ``it was an economic question. I had to develop the visualizations. They didn't want to pay for me to put all the time in we would need to fully do a good explainer'' (I8). In describing resistance to visually representing uncertainty in graphic reports, a journalism interviewee explained: ``They know they're going to have to help work on labels to describe it. That's energy that could, in their eyes, perhaps be better spent elsewhere'' (I5).

A related set of concerns framed uncertainty as \textbf{opportunity for author error}.
Multiple interviewees provided rationales for not presenting uncertainty that resembled what behavioral economists describe as a perception of ``misplaced precision''~\cite{fischhoff2012communicating,manski2018communicating} for which an author might later be blamed.
One industry interviewee, who described themself as ``probably hav[ing] more stats cred than other data vis people,'' questioned whether it was ``better to add something indicating uncertainty if you don't feel like you have a lot of confidence in those numbers'' (I6), suggesting qualitative expressions of uncertainty as a likely alternative that they and other authors might consider instead. ``There's somewhat of a worry of I'm not a hundred percent. If I'm going to include uncertainty, I want it to be really good. I place both into the sense of `there's uncertainty about uncertainty, so if I put a number on it, what if I'm wrong?''' (I6).
The possibility that not presenting uncertainty could mean viewers' would infer even more precision was not raised by authors.  


When asked for reasons for uncertainty omission that they had observed among colleagues, a journalism interviewee (I3) described how often authors found it difficult to calculate and describe uncertainty or the model assumptions used to derive it. 
When asked what they perceived to be the most helpful resource to bring about more uncertainty representation in the media, they concluded ``What I really want are better word equivalent versions.'' Other journalism and freelance interviewees echoed the challenges with explaining uncertainty (I4, I5, I12).

\begin{figure*}
\centering
\includegraphics[width=\linewidth]{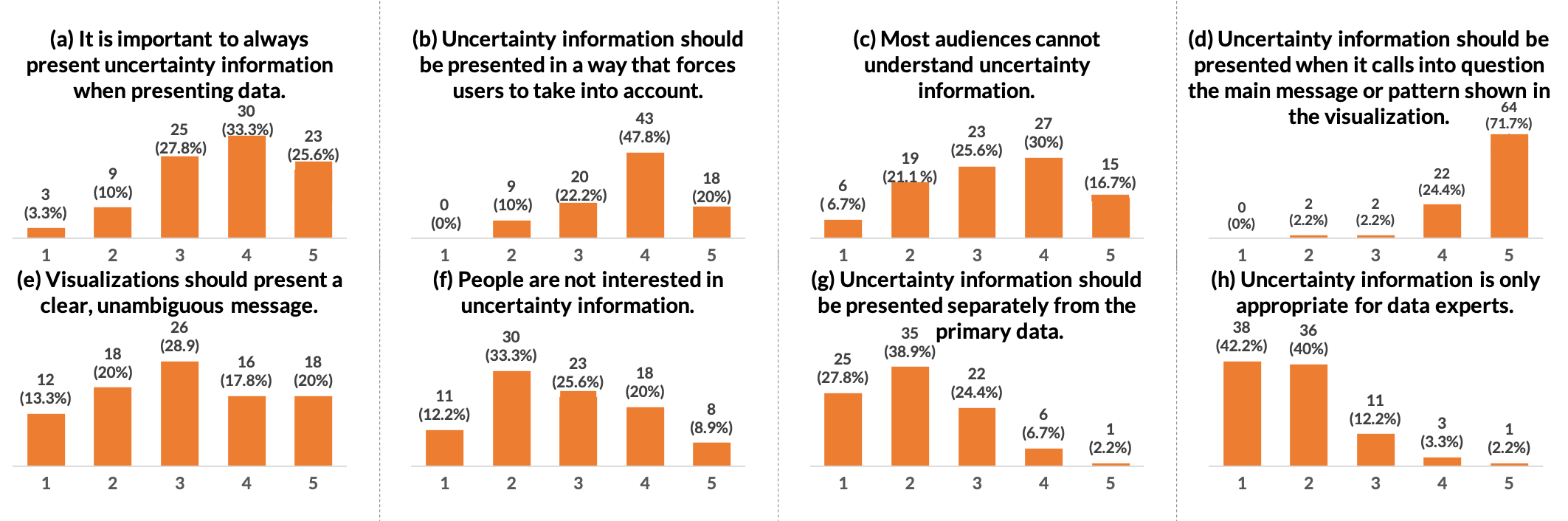}
\caption{Survey participants' ratings of their agreement with statements on a 5pt scale (1=Strongly Disagree, 5=Strongly Agree).}\label{fig:survey_result}
\end{figure*}

\subsection{Attitudes on the Importance of Uncertainty}

A majority of the surveyed authors expressed a belief that uncertainty should be visualized or explicitly represented more often than it currently is than less.
An optimism about uncertainty representation is suggested by a left skewed distributions of agreement among online survey respondents with the statements ``It is important to always present uncertainty information when presenting data'' (Figure~\ref{fig:survey_result}a), ``Uncertainty information should be presented in a way that forces users to take it into account'' (b) and ``Uncertainty information should be presented when it calls into question the main message or pattern shown in the visualization'' (Figure~\ref{fig:survey_result}d).

This sentiment was echoed by most of the interviewees. Nearly all interviewees described visualizing uncertainty more often as a goal or remarked on a feeling of responsibility that they should be doing it more often.

Some of the reasons behind the desire to visualize uncertainty more often appeared to be driven by a sense of \textbf{responsibility to accurate presentation}. 
Survey respondents insinuated the importance of transparency in presentations of data, suggesting that omitting uncertainty was misleading or even lying. One interviewee described feeling ``mildly guilty'' for not representing uncertainty more often (I6). 
Several interviewees also described how they perceived uncertainty representation to have \textbf{educational potential} (I5, I7, I13). Another journalism interviewee described wanting to show uncertainty more often to help their audiences become accustomed to what scientific measurement is like: 
``I feel like once readers start to understand that there's uncertainty inherent in a lot of these things, it's still significant. To get used to the idea that this can be very rigorous and still have a range of uncertainty, that's where I feel like I need to get better at not just saying, `Well, we'll just show these data points''' (I5). 
A survey respondent similarly argued for uncertainty presentation as a default because ``otherwise people won't get better at realizing nuances in data.'' 
These comments suggest that some authors perceive uncertainty omission as a normative practice that prevents viewers from developing certain abilities.  


\subsubsection{Inconsistency \& Heterogeneity in Attitudes}
Perhaps the most striking observation in the survey responses and interviews was the degree of heterogeneity and even inconsistency in attitudes around some considerations. Despite the left skew in agreement with ``It is important to always present uncertainty information when presenting data,'' nearly 28\% of survey respondents rated their agreement as neutral and another 14\% leaned negative. Agreement with statements that ``visualizations should have a clear, unambiguous message'' (Figure~\ref{fig:survey_result}e), that ``most audiences cannot understand uncertainty information'' (Figure~\ref{fig:survey_result}c),  and that ``people are not interested in uncertainty information'' (Figure~\ref{fig:survey_result}f) were even more ambiguous. 

Some inconsistencies may stem from the difficulty of capturing the heterogeneity of visualization design and interpretation strategies with simple high level statements. For instance, task-specific concerns are inevitable in visualization design. Several survey respondents and most interviewees described taking care to identify the task or inference that a visualization should support, and indicated that some tasks might warrant presenting uncertainty more than others. However, the tension I observed---between authors' various conceptions of the communicative function of a visualization and their associations with uncertainty---seemed unlikely to be fully explained by taking into account tasks.
Despite the optimism among most of the interviewees, in nearly every interview, I observed some form of inconsistency between the implied desire for more uncertainty and the interviewee's statements about sufficient or ideal practices for uncertainty.

For example, one journalism interviewee (I3) acknowledged the importance of uncertainty representation and described their stance that it should happen much more often, including in their team's work. However, when later asked for instances of projects where they thought uncertainty should have been more explicitly represented, they defended the practices they commonly use to ``vet'' the information for signs of unreliability before publishing the visualization. 
Another industry interviewee with statistical training in their background mused that ``in theory, I should have a lot of uncertainty. I'm working with very hardcore statisticians and machine learning experts and scientists, and yet I don't explicitly include uncertainty in my visualizations ever. It was actually quite interesting, after you prompt, thinking about why'' (I6).  
Hence authors who seem clearly capable of calculating and representing uncertainty well, who occupy roles in which they have the freedom to experiment, and who expressed interest in representing uncertainty more often, seemed unable to produce self-satisfying reasons for why they did not represent uncertainty more often. 

\section{A Rhetorical Model of Uncertainty Omission}
I propose a model to explain how authors' perceptions of the function of visualizations and uncertainty may contribute to a norm of omission despite professed acknowledgement of the value of uncertainty.
The model is rhetorical, in that it explains how authors make decisions aimed to guiding interpretations within a larger system of communication conventions and semantic associations~\cite{hullman2011}. The model is epistemological, in that it concerns how knowledge is constructed. 

The model is premised upon the fact that uncertainty omission is a norm. Three tenets capture pervasive associations that give rise to this norm. Tenet 1 states that in communicative visualization, visualizations are representations of a message or signal that an author wishes to convey to an audience. 
Tenet 2 describes how analytical process validates the signal for both authors and viewers, though only authors typically have access to this process while viewers trust the process without experiencing it.  
Tenet 3 proposes that uncertainty functions as a question or seam in the message that threatens the validity of the signal for statistical, attentional, or normative social reasons. 

\subsection{Premise: Uncertainty Omission is a Norm}
The low proportion of uncertainty visualizations out of authors' total visualizations and fact that nearly 20\% of authors did not recall communicating uncertainty at all in the last year suggests that avoiding explicit representation of uncertainty in visualization presentations is a norm, or generally accepted standard for behavior. 

Multiple comments from interviewees and survey respondents support this tenet. One journalism interviewee explicitly described increasing the frequency with which they visualized uncertainty in their work as having the potential to ``help shift what the norm is overall for all the public in terms of representing these things'' (I5). Another journalism interviewee summarized the relative lack of uncertainty communication in their organization as ``no one cares'' (I3). An industry interviewee who previously worked with scientists described how ``in scientific communication for the public, the general consensus has been don't quantify your uncertainty''  (I8). 
After commenting on one organization's use of error bars and other uncertainty representation, another freelance interviewee commented ``I do think a lot of places are mired in this cultural practice that no one's ever challenged and thought, `Do we need to provide this?' Maybe some people need to, but most people do not'' (I4). 
A survey respondent described how their work ``would not be accepted by politicians because other analysts never describe uncertainty so my more robust visualizations appear less valid if I am transparent.''

\subsection{Tenet 1: A Visualization Expresses a Signal}
Despite ambiguity in the distribution of agreement among surveyed authors with the statement ``A visualization should present a clear message,'' most interviewees and many survey respondents alluded to conveying a signal or message as the primary function of visualization-based communication. 

A journalist interviewee described how conveying a clear signal with a visualization was ``related to how you think about stories as a journalist, you want to simplify and deliver one core message with some contextualization'' (I2). Another interviewee described how hearing from their scientist collaborators that ``we helped make their point a little bit clearer to folks''(I5) was a sign of visualization effectiveness. Multiple other interviewees confirmed this (I3, I6, I9), for example, an industry interviewee describing how the best feedback on a visualization was how to make the signal stronger (I6).

The signal in a visualization was not expressed by authors as a particular kind of message or statistic, but instead appeared to represent a simplification or a proxy for an idea in which the author and the viewer share an interest.
Authors mentioned specific examples from their work at times, including the magnitude of single values or differences between values, to trends, to relationships or connections between data or trends.  

Authors also suggested that a signal is a crystallization or abstraction of something more complex. One freelance interviewee described the restraint they must consistently maintain in designing a visualization, ``There's always a creeping temptation, not just from the client to say as much as possible, but also myself. I get caught up in the mood of a subject and think, `That's interesting. That's interesting.''' (I4).

Finally, many authors implied that a signal has a truth value, as in equating uncertainty visualization with ``how to convey the possibility  of divergence from a projection without casting doubt on the fundamental measurements or reports,'' descriptions of using the design process to identify patterns that were ``more signal than noise'' (I3) and the use of terms like ``iffy'' to describe some signals (I2).

\subsection{Tenet 2: Process Validates Signal}
How does the signal described in Tenet 1 come to be seen as worth conveying by an author? 
In other words, what validates the signal in the eyes of the author, and subsequently in the eyes of the viewer? 

\subsubsection{Faith in Process Among Authors}
A faith in process was one of the most common rationales among authors for why omitting uncertainty information was acceptable. Interviewees repeatedly emphasized the important role of process in validating claims made by visualizations. For example, a journalism interviewee described how it is a journalist's job is to make sure that information they are presenting is ``not so iffy that you need [uncertainty information]'' (I2). 
Several other interviewees described how they often collaborated with scientists or data analysts who would be responsible for the statistical analysis behind a data visualization. In these scenarios, faith in the analyst's or scientist's process as assumed. For example, one interviewee alluded to how these data experts would ``make sure things are significant'' (I2). Another suggested that ``Scientists generally don't assert anything if there's a fifty-one percent chance. They wait until they're really confident'' (I8). 

These comments would seem to imply that the truth value of signal is independent of the visualization, which simply expresses that signal. However, more than half of interviewees mentioned ways in which they used the visualization process to identify and clarify the signal.
For example, multiple industry interviewees framed the design process as ``tuning'' the visualization to optimize signal strength, implying that a signal was not independent of the visualization design: ``Being able to change [the color palette] gives a way of reducing some of that type of uncertainty... 
how do I make sure that if there's a signal in the dataset, you're actually able to see it?'' (I6) 
A journalism interviewee described a process akin to looking for ``visual significance'' (I5) in which visualization is used to establish if a pattern is trustworthy. They described how in visualizing data for an upcoming story, a colleague ``realized you couldn't see the pattern. It was statistically not a very strong pattern. This isn't supporting the case at all.'' 
Several interviewees described the visualization process as characterized by looking for things that are ``basically more true'' (I3) or ``more signal than noise'' (I1, I2).  
These remarks imply that authors use a threshold to determine whether a signal is valid enough to share through visualization, and that visualization plays a role in the identification and strengthening of that signal. 

\subsubsection{Faith in Process Among Viewers}
While authors look to their own or collaborators' process to validate a signal, viewers are not privy to these processes. From a statistical standpoint, it would seem that an expression of uncertainty should be required to establish trust whenever visualized data is intended for inference. 

However, multiple interviewees and survey respondents implied that most viewers who they created visualizations for did not require specific information about process or uncertainty to trust that a signal is valid. One freelance interviewee replied sarcastically, ``it's more like the opposite'' (I13). Instead, some authors described trust as a pervasive default in visualization-based communication. 
As one industry interviewee described it ``There's a participation in trust with the system produced by the information, so wherever that comes from. Most people will trust the doctor, not necessarily because the information itself was trustworthy, but because the doctor was'' (I8). In contrast to the seemingly rational expectation that uncertainty would play a role in fostering trust, the same interviewee described how a priori trust is instead a necessary \textit{precondition} to presenting uncertainty: 
``I would say that you want trust established before you show uncertainty... 
My hypothesis would be that it [uncertainty communication] may have no effect for trust development.'' 

One industry interview described how the power of ``professional judgment'' becomes salient in the transition between exploratory to explanatory data analysis:  ``If you've gotten to the point where you're presenting this and as a professional, as a scientist you have determined that this is acceptable, and so the particulars of the uncertainty... are almost like an engineer whiteboard interview. Somebody says, `What about the uncertainty bands around this point X?' You're like, `I see that too. I'm going to respond to that''' (I11). When asked if they had seen audiences asking for uncertainty communication in data scientists' presentations, the interviewee responded ``They don't care in the short term. Uncertainty is a long-term strategic problem.'' 

Another freelance interviewee described how an organization they had worked for could not assume the standard level of trust based on the controversial nature of their business, and how this status impacted their use of uncertainty visualization: ``One of the things that they were saying is because people don't trust them, they have to go overboard with authoritative looking charts and visuals... 
If they don't show workings, if they don't show error bars and all these other things, they're in a losing battle '' (I4).  
Even if viewers are not 
always consciously aware of the process used to identify the signal in a visualization, 
that they trust the results of the author's process helps explain why conveying the ``workings'' through techniques like uncertainty visualization may often not be seen as necessary.

\subsection{Tenet 3: Uncertainty Obfuscates Signal}
If visualizations provide signal, uncertainty obfuscates that signal in the eyes of many interviewees and survey respondents. 
Multiple survey respondents and interviewees alluded to perceptions of uncertainty as separate and even ``tangential'' to visualized information. Describing the role of a journalist in providing a simplified message, a journalism interviewee described how ``Uncertainty would be peripheral to the message'' (I2). 
Another journalism interviewee used a fashion metaphor to visual representation of uncertainty to an unnecessary accessory (I3).  
An industry interviewee described how uncertainty ``doesn't feel like it's actually part of the core aspect of the data visualization'' based on their experiences as a visualization expert in a large corporation where analyses are regularly conveyed through visualization (I11).

A more negative view of uncertainty as questioning the message or signal in a visualization either came up spontaneously or was confirmed via questioning in nearly every interview. 
Beyond the 17\% of survey respondents for whom not wanting to ``make the data seem questionable'' deterred them from presenting uncertainty, multiple survey respondents also associated the term ``uncertainty visualization'' with questioning, as in 
``a visualization that asks a question,'' and 
``how to convey the possibility of divergence from a projection without casting doubt on the fundamental measurements.''

When it came to exactly how uncertainty obfuscates a signal, authors alluded to several distinct ways. The first way is best described as statistical: ``Error could be so large that it invalidates the data.'' 
When asked if their organization, a large news publication, had developed explicit guidelines or norms for communicating uncertainty, a journalism interviewee replied sarcastically that ``we only show it when we want to conclude the opposite. Uncertainty comes out most when you want the trend to not be true'' (I3). A survey respondent conversely described how viewers' ``are interested in uncertainty when it is doesn't overwhelm a clear takeaway.'' 
Another survey respondent openly admitted to choosing to omit uncertainty from a visualization because ``data wasn't reliable and uncertainty seemed too big.''


A second way is best described as attentional, suggesting the potentially distracting nature of uncertainty and the work it entails of viewers. One journalism interviewee alluded to how it can be difficult to motivate the presentation of uncertainty from the standpoint of what it provides for the viewer besides extra work: ``you're analyzing your audience. Sometimes the editor is going to say, `we're going to lose views on this', so they'll want to omit it'' (I5). Another freelancer explained the non-value of uncertainty in many instances by saying ``it's a constant battle to get bums on seats, but also then to make it worthwhile for that bum being on the seat'' (I4). 


A third way is best described as procedural. Here, uncertainty appears to operate like a seam, or ``crack'' in the visualization that signals the process behind it. One survey respondent described how uncertainty visualization was to them ``The answer to the question someone will inevitably ask, `how did you get these numbers?''' If this question is in fact inevitable, it would seem more efficient for a visualization author to try to answer the question in advance of being asked by presenting uncertainty.
However, many authors seemed motivated to avoid this question where possible, lending further evidence to a belief that viewers' trust is a priori.

One reason for avoiding references to process is that authors may not feel prepared to fully explain why the signal they were presenting was valid. That many authors found it challenging to calculate uncertainty and worried about producing wrong explanations provides some support for this possibility. 

A final way that authors implied that uncertainty can obfuscate is social. Uncertainty is again analogous to a seam, but is perceived as undermining the author's credibility based on a violation of the trust assumptions underlying norms. As one interviewee describes, ``if you go to the extent of trying to reassure people with workings and assumptions and methodologies, that sense of administration wrapped around a graphic is off-putting. It gives you a reading task. Actually, sometimes it makes you think that people are trying too hard. There must be something even more that they're not telling me about'' (I4). 

It is worth noting that many authors seemed confident in stating rationales, as though they perceived them to be truths that do not require examples to demonstrate. It is possible that rationales for omission represent ingrained beliefs more than conclusions authors have drawn from concrete experiences attempting to convey uncertainty.

\section{Modeling Inference in Communicative Visualization}
A reader may find it easy to spot inconsistencies across the model tenets. For example, the efforts that authors go through to identify and clarify a signal using visualization may seem unwarranted if viewers are thought to default to accepting the truth value of signals they are presented with. 
Or, the faith that authors have in their process to validate a signal may seem at odds with their desire to avoid presenting uncertainty when it can act as a seam that exposes aspects of that process. 
Exposing inconsistencies may help in encouraging authors to reconsider how their actions contribute to a norm of uncertainty omission despite good intentions. 

Many statements authors made suggested that pragmatic concerns about perceived consequences of communicating uncertainty often outweighed their desire to communicate it.  
As a result, many authors' might find blanket arguments that uncertainty communication is a duty or moral imperative, as some have suggested~\cite{fischhoff2012communicating}, unsatisfying. 
For example, the pragmatic author might believe that the challenges of uncertainty for viewers make it credible to omit uncertainty whenever it seems unlikely to change the inferences or decisions of the audience.

How then might proponents of uncertainty visualization convince authors that uncertainty is critical to communicate despite the many challenges with doing so? 
Reasoning more formally about 
how inferences arise with uncertainty communication versus is one step toward dismantling beliefs of omission as a defensible strategy on logical, rather than ethical, grounds.
Toward this end, 
I adapt a theory of statistical graphical inference for exploratory data analysis proposed by Gelman ~\cite{gelman2003,gelman2004exploratory} to communicative visualization.
I demonstrate through examples how, in the absence of uncertainty, a viewer's implicit specification of a model to compare to observed data is drawn from a seemingly infinite space of possible models. Uncertainty representation necessarily constrains the set of possible models. 

\subsection{A Formal Theory of Visualization-Based Inference}
Our goal is to formalize a definition of ``signal'' and of the process that produces a signal as described in the rhetorical model. 
Several statisticians have proposed that 
the process by which a person visually examines a graph to evaluate the strength of evidence is analogous to a statistical test~\cite{buja1988elements,buja1999inference,gelman2003,gelman2004exploratory,wickham2010}. While the primary goal of these analogies has been to define methods for generating visualizations to compare to observed data, 
accounts by Buja et al.~\cite{buja1988elements} and Gelman~\cite{gelman2003,gelman2004exploratory} also acknowledge the role visualizations play in informal statistical inference. In this process, a visualization is compared to an \textit{implicit} (i.e., imagined) reference distribution to determine whether the visualization presents something interesting.  

\begin{figure}[htb]
 \centering
  \includegraphics[width=\columnwidth]{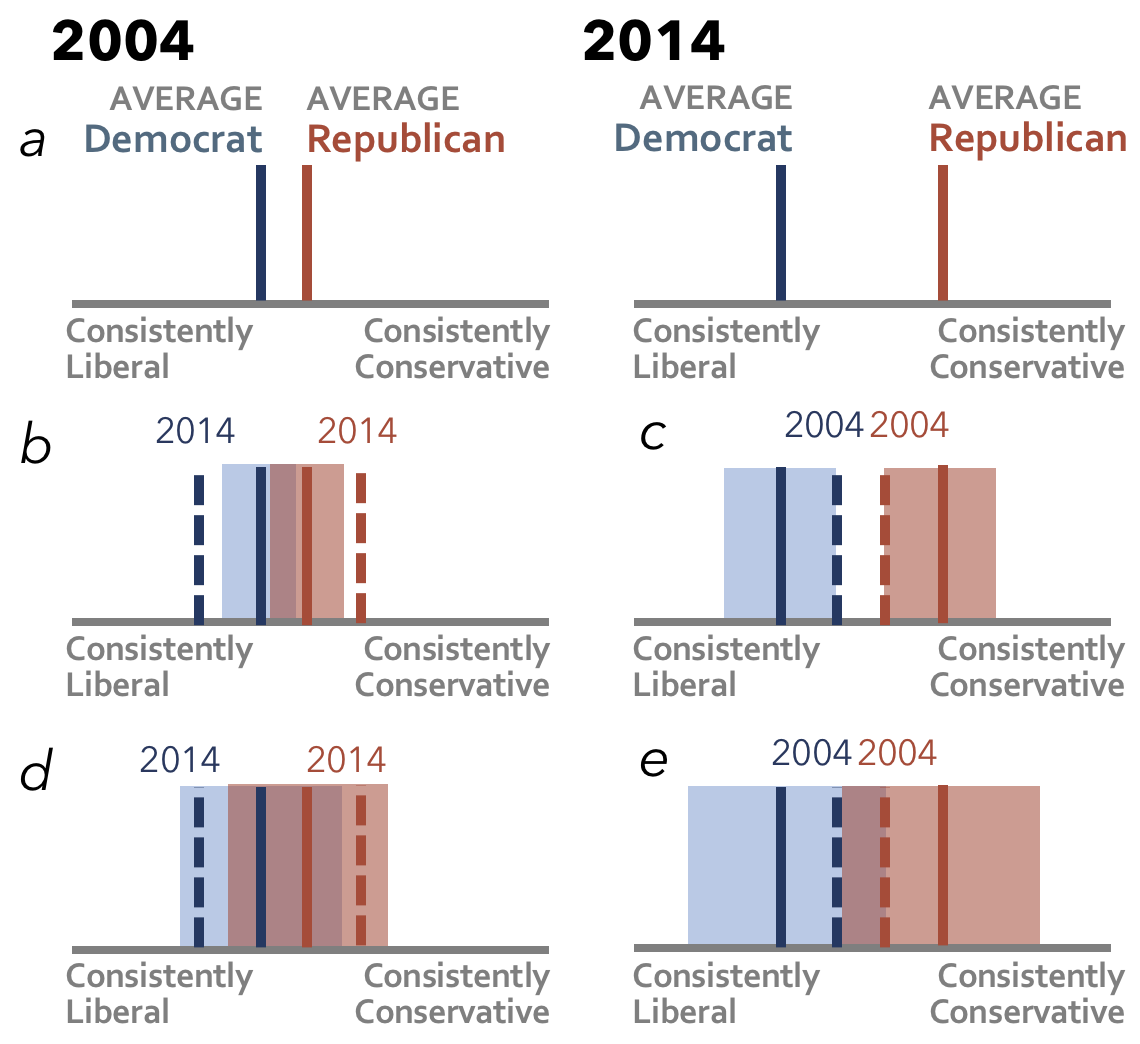}
  \vspace{-0.25in}
  \caption{An adaption of a visualization presented by the Pew Foundation~\cite{pew}, top, which shows Democratic and Republican respondents' mean scores in 2004 and 2014 on a survey that measures consistency in political attitude. The center of each $x$-axis represents political neutral answers. Shifts in the means across the two years indicate changes in how consistent respondents political attitudes are. Parts b, c, d, and e depict varying model specifications that a viewer might formulate to judge the strength of the visualization's signal: assumed variance is lower in b and c than d and e, and b and d fit a model to 2004 data versus 2014 data in c and e. }
  \vspace{-0.1in}
  \label{fig:pew}
\end{figure}

Gelman's proposal~\cite{gelman2003,gelman2004exploratory} describes the examination of a visualization is a ``model check''\footnote{Visual methods for enabling confirmatory analysis by hiding observed data in a set of plots sampled from a ``null plot distribution''~\cite{buja2009}, also known as LineUps~\cite{beecham2016,majumder2013,wickham2010}, can be considered one type of model check, though model checking can also involve inferences aimed at understanding \textit{how} data deviates from model predictions.}
Assume a person looking at a visualization to estimate some parameter values of interest $\theta$. 
To take a simple example, imagine a viewer looking at Figure~\ref{fig:pew}a, which summarizes the consistency of  political attitude among samples of respondents who identify as Democratic versus Republican over two time periods. Here $\theta$ might represent a vector of estimates of the true Democrat and Republican averages in 2004 and 2014. 

We can imagine a viewer judging the strength of a signal in Figure~\ref{fig:pew}a via an implicit (i.e., mental) comparison of the visualized data $y$ to a reference distribution $T(y^{rep})$ (i.e., a distribution of outcomes produced by a hypothesized data-generating model). We can liken this to comparing  standardized test statistics $T(y)$ and $T(y^{rep})$.
In a Bayesian framework, the data generating process is the posterior predictive distribution $p(y^{rep}|y)$, or the distribution of the outcome variable implied by a model that used the observed data $y$ to update a prior distribution of beliefs about the unknown model parameters $\theta$~\cite{gabry2019}. A viewer compares imagined simulations from $p(y^{rep}|y)$ to the observed data, looking for discrepancies. A discrepancy function $D(y, y^{rep})$ can be thought of as a measure of the deviation between $T(y)$ and $T(y^{rep})$. 
As commonly assumed in signal detection theory~\cite{green1966}, the viewer might subject the result of this function to some individualized threshold (i.e., a certain perceived probability that the value must be below) to decide if the visualization presents ``something interesting,'' similar to how an $\alpha$ level is used in assess a $p$-value in NHST.  

What reference distribution might a viewer assume in examining Figure~\ref{fig:pew}a?
One possibility is that a viewer is conducting a ``visual hypothesis test.'' $T(r^{rep})$ is drawn from a ``null plot distribution''~\cite{buja2009} specifying plots that might be seen if there were in fact no difference between the Democrat and Republican data.
Here, the discrepancy function $D(y_{2014}, y^{rep}_{2004})$ could produce an intuitive estimates of the probability that the $T(y_{2014})$ could have in fact been generated by $p(y^{rep}_{2004}|y_{2004})$, similar to a $p$-value~\cite{gelman2003}. 

Conceiving of graphical inference as a model check is not to suggest that viewers enact these processes consciously or identically to the analogous statistical processes. In fact it may be more realistic to assume that viewers may rely on various approximations to estimate $p(y^{rep}|y)$ and $D(y, y^{rep})$ (e.g,~\cite{khaw2017risk,kim2019}). The value of the formal model lies in the way it summarizes potential mechanisms of graphical inference to enable principle reasoning about the impacts of different visualizations---in particular, the implications of omitting versus visualizing uncertainty.   



The posterior predictive distribution $p(y^{rep}|y)$ is a function of a viewer's prior beliefs about parameter values, their model specification, and their perception of the data $y$ including its variance. 
In the absence of uncertainty representation, it seems unlikely that viewers assume a common level of uncertainty in the observed data and use it identically in fitting the model\footnote{A viewer might fit $p(y^{rep}|y)$ using only the observed values of predictors or new observations of the predictors~\cite{gabry2019}}. For example, viewers of Figure~\ref{fig:pew} top are free to assume any values for the variance in each expected distribution of means (though distributions that exceed the $x$-axis range would not be congruent with the meaning of the bounded measurement). Figure~\ref{fig:pew}b and c, versus d and e, demonstrate two possible assumptions of variance (where the two parties' variance is assumed to be equal). These are just two variations on many different ways the viewer might mentally ``fill in'' the missing information about the authors' intended inferences to decide if a chart presents a valid signal. 

   
For Figure~\ref{fig:pew}, as in many statistical applications, there are various other degrees of freedom in the viewer's specification of $p(y^{rep}|y)$ which might return different results.
Comparing Figure~\ref{fig:pew}b and c (or d and e) demonstrates how two applications of the basic simple model structure, with a change to only which year's data is used to fit $p(y^{rep}|y)$, can lead to different conclusions by the viewer.
Alternatively, a viewer might imagine both means in each year as samples from a Gaussian distribution centered at political neutrality to estimate the probability that one or both of the years represent real differences, and compare how large the discrepancies from the model are across years.

\begin{figure}[htb]
    \includegraphics[width=0.5\textwidth]{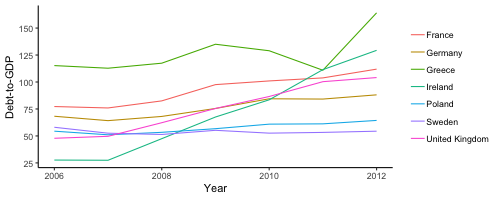}
   \vspace{-0.26in}
    \caption{A visualization comparing government debt across countries. A viewer might perceive a disparity between the average increase in debt for Ireland between 2006 and 2012 and that of the other countries. Changing the visualization by removing data (e.g., Greece) or extending axes ranges (e.g. $y$-axis to 0) could impact the strength of signal that the viewer concludes.}
  \label{fig:debt_to_gdp}
\end{figure}

As another example, consider Figure~\ref{fig:debt_to_gdp},
a line chart based on a real world example of estimates of some countries' debt to GDP ratios~\cite{oecd}. 
Assume that the author presents the visualization with text indicating the intended message, like the title ``Ireland's relative debt grew quickly between 2006 and 2012.'' Even with information about the intended signal, there are multiple models that a viewer might use to determine how much the trend for Ireland differs from that of the other countries. If we assume a viewer is interested in the high level disparity in the slope of Ireland from 2006 to 2012, then a simple linear model structure $y$ = $\alpha$ + $\beta$*$x$ could suffice, but again the question arises of what data is used to fit the model. One possibility is that the viewer imagines a distribution of replicates ($T(y^{rep})$) drawn from a predictive distribution fit to the data from 2006 and 2012 from all countries other than Ireland. However, should the viewer assume the presented countries are intended to represent the larger set of all European countries other than Ireland, or rely only on those in the visualization? With no presentation of uncertainty, it is difficult to say which better captures the authors' intention. Viewers may also bring different prior beliefs; imagine how the $p(y^{rep}|y)$ of a viewer who expects that Ireland underreported their national debt leading up to a financial 2009 crisis might differ from that of a viewer with an uninformed prior. 



\subsubsection{Contradictions Exposed by the Model}
From the rhetorical model, it seems likely that authors believe viewers' model specifications, discrepancy functions, and thresholds will be very similar to if not the same as theirs, as this is the ``machinery'' that defines the signal for the author. 
However, that multiple authors described using the visualization ($T(y)$) to judge the significance of a pattern also implies that the data generating process that produces the imagined $T(y_{rep})$ ($p(y^{rep}|y)$ depends on the particulars of the visualization $T(y)$. For example, without seeing the countries shown in Figure~\ref{fig:debt_to_gdp}, it would seem difficult to assume a viewer with a $T(y_{rep}$ already in mind.

The discrepancy function $D(y, y^{rep})$ will naturally change with $T(y)$ as well. For example, if the author of Figure~\ref{fig:debt_to_gdp} were to remove a line from the reference set of countries that deviates somewhat from the rest (e.g., Greece), then $D(y, y^{rep})$ will presumably also change. Similarly, increasing the range of values on the $y$-axis might imply to some viewers that it would be reasonable to expect other countries to have higher intercepts or greater slopes than those visualized.  
An author's perception that a signal has a truth value, which is defined outside of the final communicative visualization (e.g., through significance testing), is therefore inconsistent with the graphical inference model.
These and other contradictions highlight the value of a formal model for systematically demonstrating the implications of omitting uncertainty. 


\subsubsection{How Uncertainty Can Help}
With a formal model in place, it becomes possible to ask, What role does visualizing uncertainty play in the graphical inference process? Just as viewers may assume different reference distributions in determining the strength of a visualized signal, the role that explicit representation of uncertainty plays in the inference process can vary based on how uncertainty is formulated.

Ideally, there is a correspondence between how an author specifies uncertainty and the graphical inference process that gave rise to their judgments of signal strength. For example, imagine a visualization $T(y)$ that plots a new observation (or set of observations, as in Figure~\ref{fig:pew}b that one wishes to assess in light of an explicitly visualized reference distribution. The intervals surrounding the 2004 values in Figure~\ref{fig:pew}b essentially convey $T(y_{rep})$.  
Compared to a mean value with no interval, 
visualizing uncertainty around each estimates facilitates the intended inference by greatly reducing degrees of freedom in the viewer's graphical inference process.

However, assuming a correspondence between the author's inference and presented uncertainty is dangerous for several reasons. If the author of a visualization like Figure~\ref{fig:pew}b were to have also taken care to visualize measurement error around the 2014 values, then it becomes more difficult for viewers to be sure that the reference distribution they infer was that which the author intended without further clarification by the author. 
Perhaps more likely, based on the difficult of calculating and visualizing uncertainty, is that authors visualize uncertainty in whatever way seems convenient rather than reflecting on how to convey their own assumed reference distribution. 
Some authors' professed challenges communicating uncertainty of any kind suggest that 
even when a reference distribution is intuitively salient, 
it may not be obvious how to 
produce a depiction specifying $p(y^{rep}|y)$ for viewers. 
Finally, research on the challenges many viewers face interpreting expressions statistical constructs like standard error or confidence interval error bars~\cite{belia2005researchers,hoekstra2014robust} suggests in is unlikely that all audiences could reliably infer the author's modeling assumptions from 
some uncertainty visualizations without additional description to guide their inference.


Given that an uncertainty visualization does not necessarily \textit{ensure} that all viewers formulate $T(y^{rep})$, $p(y^{rep}|y)$, or even $T(y)$ in the same way, some authors may feel validated in their choice to omit uncertainty. 
Perhaps it may seem permissable to omit uncertainty if one's analysis process has passed some sufficiently high level of robustness or rigor, out of a desire to avoid potential heterogeneity in beliefs that might result from visualizing it.
However, an author would need to believe that the diversity of forms $T(y^{rep})$ takes with uncertainty is even greater than that taken without. Visualizing uncertainty conveys more information about potential reference distributions, and consequently data generating processes, 
even when it does not directly correspond to the author's graphical inference process. This reduces the amount of information that a viewer must mentally ``fill in,'' which would seem to necessarily reduce variance in interpretations across viewers.    
Instead of throwing up one's hands at potential heterogeneity in uncertainty interpretations, one might conclude that authors should not only strive to depict uncertainty more, they should strive to depict uncertainty as defined by their own inference process. One might call this process ``reference model visualization'' to suggest its importance for guiding viewers' unavoidable implicit inferences. 

\section{Discussion}

The results presented here suggest that uncertainty communication is far from a solved problem for visualization authors. Various rationales for omitting uncertainty emerged from my survey and interviews. However, many authors also said they would like to communicate uncertainty more frequently, and expressed intuitions that doing so 
would shift people's general tolerance for and ability to reason with uncertainty. 
This tension and others lead me to suggest that, as one interviewee put it: ``uncertainty is a long-term strategic problem'' (I11).

The rhetorical and graphical inference models provide a theoretical basis for reasoning about uncertainty omission, one that is consistent with authors' general perceptions of uncertainty visualization as ``a visualization that doesn't have a clear message.''
These contributions are a call to action for visualization and human computer interaction researchers to extend their efforts beyond the creation of specific techniques and applications, and to consider the larger background of incentives, norms, experiences, and insecurities that undergird uncertainty omission. Are we in a ``bad equilibrium''~\cite{manski2018lure} as one economist has asked? If so, what will it take to change this state? In an era of increased data rhetoric, political propaganda, and public fears around predictive modeling, research efforts toward shifting communication norms to require uncertainty communication are more critical than ever. 

The importance of understanding the problem of uncertainty omission at scale is not to say that researchers should lessen their efforts toward producing widely accessible and generalizable techniques for visualizing uncertainty. One interviewee described how an engineer at their company argued, ``Uncertainty visualization is solved. Just use these hypothetical outcome plots.'' The interviewee mused ``That's great, except people aren't using them.'' That many authors described challenges and insecurities about visualizing uncertainty suggests that visualization researchers may be failing in disseminating their results in terms that meet authors where they are at. Tools that help visualization authors not just visualize but calculate and explain uncertainty are well motivated by my results, as is research into the value that authors perceive in more approximate uncertainty communications, like showing views that problematize a default level of aggregation or using perceptual effects to convey imprecision.


Arguments that communicating uncertainty is an author's moral imperative~\cite{fischhoff2012communicating} are one way researchers and authors may try to inspire changes in practice. 
Indeed, some authors I surveyed and interviewed expressed a belief that uncertainty communication was a responsibility. However, even these authors admitted to not visualizing uncertainty as often as they thought they should (or at all). 
While readers may find it ethically questionable that some authors would omit presenting uncertainty when it could call into question the message of the graphic, I intentionally avoided such value judgments in analyzing and compiling the perceptions I reported on. 
Instead, the formal model I presented aims to provide logical ground for countering pervasive and often pragmatic rationales for omission. 

Of course, an abstract logical argument may not be of any more interest to authors than an ethical one. However, it may be possible to help authors consider the implications of uncertainty omission versus inclusion through concrete examples. For example, researchers could develop tools that allow an author to simulate possible reference distributions that viewers may adopt. 
Another promising approach to making the implications of omission versus inclusion more concrete is to apply decision theory to help readers reason about the impacts of their visualization. Such an approach could motivate design tools that prompt an author to consider possible ``worst case'' decisions (or actions) viewers might take under uncertainty omission versus inclusion. The latter approach may be particularly useful in scenarios where the author would otherwise not think through 
how their visualization might impact viewers' decisions or beliefs. Finally, recent approaches to visualization evaluation that apply Bayesian cognition to evaluate how viewers update their beliefs after using a visualization~\cite{kim2019} may be a further avenue for helping authors perceive the implications of uncertainty more concretely. 


Beyond reasoning about uncertainty omission, formalizing ``intuitive'' statistical graphical inference could increase awareness among researchers and practitioners around the cognitive and perceptual processes that give rise to perceived signal in a communicative visualization. A formal science of visual communication stands to improve visualization-practice writ broadly, while integrating uncertainty as part of the definition of a visualization. Researchers could, for example, develop tools to help authors articulate the graphical inferences that they target and create appropriate expressions of uncertainty to communicate these to viewers.   

\subsubsection{Limitations}
I surveyed and interviewed 103 visualization authors. The beliefs I summarized may not describe all authors' impressions. In particular, I observed considerable heterogeneity in authors' practices and perceptions around the value of uncertainty communication. That a policy of always including uncertainty information in a visualization could seem obvious to some authors and baffling to others tells of the complexity of this topic in communicative visualization.

As described earlier, the participants that I surveyed may over-represent authors who feel that uncertainty visualization is important. 
The set of influencers I interviewed may overrepresent authors who are sympathetic or at least open-minded enough about uncertainty to agree to an interview.  
Interview and survey results may also be biased by participants' abilities to retrospectively recall their choices and rationales around uncertainty communication.

\section{Conclusion}

Most visualizations outside of scientific journals do not explicitly represent uncertainty information. This work asked, why? By surveying and interviewing authors who regularly create visualizations for others, I identified perceptions, practices, challenges, and attitudes associated with uncertainty visualization. My results suggest that many authors are optimistic about the importance of visualizing uncertainty, but may face challenges calculating, visualizing, and explaining uncertainty to viewers. However, these challenges were not sufficient to explain why authors who acknowledge the benefits of uncertainty might default to omitting uncertainty. To summarize how a norm of uncertainty omission might seem reasonable, I contribute a rhetorical model that describes pervasive associations and expectations about uncertainty among authors. I apply a formal model of of graphical statistical inference to shed light on how visualizations are examined to determine the strength of a signal, and demonstrate how uncertainty reduces (though does not necessarily eliminate) degrees of freedom in viewers' inferences.

\section{Acknowledgements}
I am grateful to the survey respondents and visualization experts provided their feedback. Thank you to Yea Seul Kim and Alex Kale for comments on a draft.


\bibliographystyle{abbrv}
\bibliography{valueuncertainty}

\end{document}